\def\Title{\Huge Mathematical foundations of quantum information:
Measurement and foundations\thanks{
Originally published in Japanese as: Masanao Ozawa, Mathematical Foundations of Quantum Information (in Japanese), Sugaku 61 (2), 113--132 (2009); doi:10.11429/sugaku.0612113.
Based on a plenary address at the 2008 Mathematical Society of Japan Autumn Meeting, Tokyo, Japan, September 25, 2008. 
}}
\def\Author{Masanao Ozawa}

\documentclass[12pt]{article}
\evensidemargin=\oddsidemargin
\usepackage{amsmath,amssymb,amsthm,graphicx,color,bm}
  \let\Huge=\Large

  \let\Large=\normalsize
  
\newtheorem{theorem}{Theorem}

\theoremstyle{definition}

\theoremstyle{remark}

\theoremstyle{axiom}
\newtheorem{axiom}{Axiom}
\theoremstyle{postulate}
\newtheorem{postulate}{Postulate}
\numberwithin{equation}{section}
\renewcommand{\cal}{\mathcal}

  \newcommand{\beq}{\begin{equation}}
  \newcommand{\eeq}{\end{equation}}
  \newcommand{\beql}[1]{\begin{equation}\label{eq:#1}}

  \newcommand{\beqa}{\begin{eqnarray}}
  \newcommand{\eeqa}{\end{eqnarray}}
  \newcommand{\beqas}{\begin{eqnarray*}}
  \newcommand{\eeqas}{\end{eqnarray*}}
  
\newcommand{\C}{{\bf C}}

  \newcommand{\Q}{{\bf Q}}
\newcommand{\R}{{\bf R}}
  \newcommand{\V}{{\bf V}}

  \newcommand{\al}{\alpha}
  \newcommand{\be}{\beta}
  
  \newcommand{\da}{\dagger}

  \newcommand{\ep}{\epsilon}
  \newcommand{\et}{\eta}

 \newcommand{\id}{{\rm id}}
  
  \newcommand{\la}{\lambda}
  
  \newcommand{\nn}{\nonumber}
  \newcommand{\om}{\omega}
  \newcommand{\ph}{\phi}
 \newcommand{\ps}{\psi}
  \newcommand{\rh}{\rho}
  \newcommand{\si}{\sigma}
  \newcommand{\ta}{\tau}

  \newcommand{\sic}{\si c}
  
  \newcommand{\tc}{\tau c}
  
  \newcommand{\De}{\Delta}
  \newcommand{\Ga}{\Gamma}

\renewcommand{\And}{\wedge}

  \newcommand{\Ex}{{\rm Ex}}
  \newcommand{\Eq}[1]{Eq.~(\ref{eq:#1})}

  \newcommand{\Not}{\neg}
  \newcommand{\Or}{\vee}

  \newcommand{\Then}{\Rightarrow}
  \newcommand{\Tr}{\mbox{\rm Tr}}
  
  \newcommand{\VB}{\V^{(\B)}}

\renewcommand{\V}{V}

  \newcommand{\beqan}{\begin{eqnarray*}}
  \newcommand{\beqar}[1]{\begin{equation}\label{#1}\begin{array}{l}}
  \newcommand{\bx}{{\bf x}}
  \newcommand{\by}{{\bf y}}

  \newcommand{\dom}{\mbox{\rm dom}}
  
  \newcommand{\eeqar}{\end{array}\end{equation}}
  \newcommand{\eq}[1]{(\ref{eq:#1})}

  \newcommand{\hp}{\hat{p}}

\newcommand{\bra}[1]{\langle#1|}
\newcommand{\ket}[1]{|#1\rangle}
\newcommand{\ketbra}[1]{\ket{#1}\bra{#1}}

\newcommand{\bracket}[1]{\langle#1\rangle}

\newcommand{\bmat}{\left[\begin{array}{rr}}
\newcommand{\emat}{\end{array}\right]}
\newcommand{\bvec}{\left[\begin{array}{r}}
\newcommand{\evec}{\end{array}\right]}
  \newcommand{\bA}{{\bf A}}

  \newcommand{\bP}{{\bf P}}
  \newcommand{\bR}{{\bf R}}
  \newcommand{\bS}{{\bf S}}

  \newcommand{\bX}{{\cal I}}

  \newcommand{\cA}{{\cal A}}
  \newcommand{\cB}{{\cal B}}

  \newcommand{\cE}{{\cal E}}

  \newcommand{\cI}{{\cal I}}
  
    \newcommand{\cH}{{\cal H}}
  \newcommand{\cK}{{\cal K}}
  \newcommand{\cL}{{\cal L}}

  \newcommand{\cQ}{{\cal Q}}
  \newcommand{\cR}{{\cal R}}
  \newcommand{\cS}{{\cal S}}

  \newcommand{\tA}{\tilde{A}}
  \newcommand{\tB}{\tilde{B}}

 \newcommand{\hx}{\hat{x}}

\renewcommand{\VB}{\V^{(\cB)}}

\renewcommand{\inf}{\bigwedge}
\renewcommand{\sup}{\bigvee}
\renewcommand{\Then}{\rightarrow}

\newcommand{\val}[1]{[\![#1]\!]}

\newcommand{\cuniv}{\underline{\Or}}

  \title{{\bf \Title}}
  \author{\sc\Author \\
  \it\small Graduate School of Information Science\\
\it\small Nagoya University, Chikusa-ku, Nagoya,  464-8601, Japan}
  \date{}
  
\begin{document}
  
  \maketitle

\begin{abstract}
The purpose of this paper is to survey some topics on mathematical foundations 
of quantum information developed mainly by the present author and
co-workers for the last three decades. 
The topics include an axiomatic construction of quantum measurement theory
based on completely positive map-valued measures, 
a universally valid new formulation of the uncertainty principle
for error and disturbance in 	quantum measurements,
the Wigner-Araki-Yanase limit of quantum measurements, 
the accuracy limit of quantum computing based on conservation laws, 
and a quantum interpretation based on quantum set theory.

\end{abstract}
\section{Introduction}
Quantum mechanics was discovered in the beginning of the 20th century and 
has revealed that nature is ruled by the quantum state with peculiar uncertainty.
Various paradoxes including Schr\"{o}dinger's cat and the Einstein-Podolsky-Rosen (EPR) 
paradox were derived from the basic formalism of quantum mechanics and yet challenged
our conventional views. Nevertheless, quantum mechanics brought marvelous success
in describing and predicting phenomena originated in the microscopic world, 
and has produced a huge field of electronics technology in the latter half of the 20th century.

It is the discovery of lasers in 1960 that opened a way of controlling the quantum 
state that was a mere hypothesis about the microscopic world
to explain experiences such as the stability of atoms.
Quantum mechanics started to play a new role in describing the limitation
of our ability to control the external world.
Moreover, unconditionally secure quantum cryptography has 
recently been developed based on the idea of precisely describing 
the limitation of an eavesdropper's ability.
This new aspect of quantum mechanics emerges as a new research field
called quantum information, which has a close connection to information
science.
Since Shor  \cite{Sho94} discovered an algorithm efficiently solving 
prime factorization 
by quantum computers, the research on quantum information  
has made great progress and has produced various proposals for
application to quantum information technology including
quantum computing and quantum cryptography.

The purpose of this paper is to survey mathematical foundations of
quantum information.
In particular, we discuss the most foundational aspect of quantum
information centered at quantum measurement theory.
It should be emphasized that the new framework of quantum information 
has solved not only technological problems relative to computing and 
communication, but also several problems on foundations of quantum 
mechanics, which have been left unsolved since the emergence of quantum 
mechanics in the 1920s, and we focus more on the latter aspect of 
quantum information research.

In Section 2, we discuss quantum measurement, one of the most fundamental
notions in quantum information.
Von Neumann's axiomatization \cite{vN32} of quantum mechanics has answered
what mathematically represent quantum states and quantum observables, but 
left unanswered what mathematically represent quantum measurements.
In the 1970s, a new mathematical theory emerged about such notions
as probability operator-valued measures (POVMs), operations,
and instruments for describing various aspects of quantum measurements, 
and the problem of mathematical characterization of the notion of quantum
measurement was completely solved based on those notions \cite{83CR,84QC}. 
This theory is now an indispensable part of quantum information theory.
In Section 3, we discuss the uncertainty principle.
In 1927, Heisenberg introduced the uncertainty principle describing
the inevitable amount of disturbance caused by the back action of a 
measurement and setting a  limitation for simultaneous measurements 
of non-commuting observables.
However, his quantitative relation has been revealed not to be universally 
valid \cite{02KB5E},
through the debate on the standard quantum limit for gravitational wave detection
induced by measurement back action \cite{88MS,89RS}.
We discuss the above-mentioned debate and a new universally valid
formulation of the uncertainty principle obtained recently 
\cite{03HUR,03UPQ,03UVR,04URN}.
In Section 4, we give an outline of the recent study of the accuracy
limits for quantum computation.
This result was obtained by quantitatively generalizing
the Wigner-Araki-Yanase theorem on the limitation of 
measurement under conservation laws by using the 
new universally valid uncertainty principle above \cite{02CLU,02CQC,03UPQ}.
In Section 5, we outline the recent investigation on interpretation
of quantum mechanics.
We shall discuss simultaneous measurability of non-commuting
observables based on the new uncertainty principle \cite{06QPC,07SMN}
and a new interpretation of quantum mechanics 
based on quantum set theory \cite{07TPQ}.

\section{Quantum measurement theory}
\subsection{Axioms for quantum mechanics}

Axioms for quantum mechanics due to von Neumann \cite{vN32} are formulated as follows.

\begin{axiom}[Axiom for states and observables]
To every quantum system $\bS$ there is a uniquely associated 
Hilbert space ${\cal H}$ called the {\em state space} of $\bS$. {\em States} of $\bS$ are represented by density
operators, positive operators with unit trace on $\cH$ and {\em observables} 
of $\bS$ are represented by self-adjoint operators on $\cH$.
\end{axiom}

We follow the convention that the inner product on a Hilbert space 
is linear in the first variable and conjugate linear in the second.
A state of the form $\rh=\ketbra{\ps}$ is called a {\em pure state} with a {\em state 
vector} $\ps$, where
the operator  $\ket{\xi}\bra{\et}$ with $\xi, \et\in\cH$ is defined by
$\ket{\xi}\bra{\et}\ps=\bracket{\et,\ps}\xi$. 
We denote by $\cS(\cH)$ the space of density operators on $\cH$.
In this paper, we further assume that {\it every self-adjoint operator on $\cH$ has
a corresponding observable of $\bS$}; the resulting theory is often called 
a non-relativistic quantum mechanics without superselection rules.

In what follows, we denote by $\cB(\bR)$ the set of Borel subsets of $\bR$
and by $E^{A}$ the spectral measure of a self-adjoint operator $A$.

\begin{axiom}[Born statistical formula]
\label{ax:BSF}
If an observable $A$ is measured in a state $\rh$, the probability distribution
of the outcome $\bx$ is given by
\beqa\label{eq:Born}
\Pr\{\bx\in\De\|\rh\}=\Tr[E^{A}(\De)\rh],
\eeqa
where $\De\in\cB(\bR)$.
 \end{axiom}
 
From the above axiom, if $A\rh$ is a trace-class operator,
the expectation value is given by $\Ex[A\|\rh]=\Tr[A\rh]$,
and if $A\sqrt{\rh}$ is a Hilbert-Schmidt-class operator,
the standard deviation is given by $\si(A\|\rh)^2=\Tr[(A\sqrt{\rh})^2]-\Tr[A\rh]^2$.
Henceforth,  $\tc(\cH)$ will denote the space of trace-class operators on $\cH$
and $\sic(\cH)$ will denote the space of Hilbert-Schmidt-class operators.

In what follows, $\hbar$ denotes the value of the Planck constant in the unit system 
under consideration divided by $2\pi$. 

\begin{axiom}[Axiom of time evolution]
Suppose that a system $\bS$ is an isolated system with the Hamiltonian 
$H$ from time $t$ to $t+\ta$.
If $\bS$ is in a state  $\rh(t)$ at time $t$, then $\bS$ is in the
state $\rh(t+\ta)$ at time $t+\ta$ such that
\beqa
\rh(t+\ta)=e^{-iH\ta/\hbar}\rh(t)e^{iH\ta/\hbar}.
\eeqa
\end{axiom}

\begin{axiom}[Axiom of composition]
The state space of the composite system $\bS=\bS_{1}+\bS_{2}$, consisting
of the system $\bS_{1}$ with the state space $\cH_{1}$ and 
system $\bS_{2}$ with the state space $\cH_{2}$, is 
the tensor product  $\cH_{1}\otimes\cH_{2}$.
Every observable $A$ of $\bS_{1}$ is identified with the observable 
$A\otimes 1$ of $\bS$ and every observable $B$ of $\bS_{2}$ is identified 
with the observable $1\otimes B$ of $\bS$. 
\end{axiom}

\subsection{Von Neumann-L\"{u}ders projection postulate}

Under the above axioms, we can make a probabilisic prediction of the result of 
a future measurement from knowing the past state.
However, such a prediction applies only to one measurement in the future.
If we make many measurements successively, we need another axiom
to determine the state after the measurement.
In the conventional approach, the following hypothesis has been proposed
\cite{vN32,Lud51}． 
 \begin{postulate}[Von Neumann-L\"{u}ders projection postulate]
Under the condition that a measurement of an observable $A$ in a state 
$\rh$ leads to the outcome $\bx=x$, the state  $\rh_{\{\bx=x\}}$  just after the measurement is given by
\beqa
\rh_{\{\bx=x\}}=\frac{E^{A}(\{x\})\rh E^{A}(\{x\})}{\Tr[E^{A}(\{x\})\rh]}.
\eeqa
\end{postulate}

In order to find the state change caused by a measurement,
von Neumann used a feature of the Compton-Simons experiment 
\cite[pages 212--214]{vN55} and generalized it to pose
the {\it repeatability hypothesis} \cite[page 335]{vN55}.

\begin{postulate}[Repeatability hypothesis]
If the physical quantity is measured twice in succession 
in a system, then we get the same value each time.
\end{postulate}

Then, from the repeatability hypothesis, von Neumann showed that 
the state change caused by a measurement of an observable 
with non-degenerate discrete spectrum satisfies the 
von Neumann-L\"{u}ers projection postulate (projection postulate, for short).
While von Neumann showed that if the spectrum is degenerate, the repeatability
hypothesis is not sufficient to determine the state change uniquely,
L\"{u}ders  \cite{Lud51} later introduced the projection postulate as the state 
change caused by a sort of canonical measurement.
 
It is well known that there are many ways to measure the same observable
that do not satisfy the projection postulate.  Thus, this hypothesis is not
taken to be a universal axiom but a defining condition for a class of measurement.
 We say that a measurement is {\it projective} if it satisfies the projection postulate.

For any sequence of projective measurements, we can determine
the joint probability distribution of the outcomes of measurements \cite{Wig63+}.

\begin{theorem}[Wigner's formula]
Let $A_{1},\ldots,A_{n}$ be observables with a discrete spectrum
of a system $\bS$ in a state $\rh$ at time 0.
If one carries out projective measurements 
of observables $A_{1},\ldots,A_{n}$ at times 
$(0<)t_{1}<\cdots<t_{n}$ and otherwise leaves the system $\bS$ isolated
with the Hamiltonian $H$,
then the joint probability distribution of the outcomes 
$\bx_{1},\ldots,\bx_{n}$ of those measurements is given by
\beqa
\lefteqn{\Pr\{\bx_{1}=x_{1},\ldots,\bx_{n}= x_{n}\|\rh\}}\quad\nn\\
&=&
\Tr[E^{A_{n}}(\{x_{n}\})\cdots U(t_{2}-t_{1})E^{A_{1}}(\{x_{1}\})U(t_{1})\rh
\\
& &\mbox{ }\times
U(t_{1})^{\da}E^{A_{1}}(\{x_{1}\})U(t_{2}-t_{1})^{\da}
\cdots E^{A_{n}}(\{x_{n}\})],\nn
\eeqa
where $U(t)=e^{-iHt/\hbar}$.
\end{theorem}

The projection postulate can be applied to a restricted class of
measurements, and has the following problems if we take it to be a
basis of quantum mechanics.
\begin{enumerate}

\item[(i)] The projection postulate cannot be applied to observables with
 continuous spectrum \cite{84QC,85CA}．

 \item[(ii)] There exist commonly used measurements of discrete observables,
 such as photon counting, that do not satisfy the projection postulate  \cite{IUO90}. 
 
\item[(iii)] There is a useful class of measurements that cannot be represented
by observables but by the so-called POVMs (probability operator valued 
measures) \cite{Hel76}.
\end{enumerate}

\subsection{Davies-Lewis instruments}

State changes induced by measurements have been called {\it quantum
state reductions}  and are considered one of the most difficult notions
in quantum mechanics.
In order to apply quantum mechanics to the system to be measured
sequentially, we need to mathematically characterize all the possible
state changes induced by measurements.

If we are given the probability distribution $\Pr\{\bx=x\|\rh\}$
of the outcome and the quantum state reduction $\rh\mapsto \rh_{\{\bx=x\}}$,
the ``non-selective'' state change caused by this measurement
is given by
\beq
\rh\mapsto T(\rh)=
\sum_{x\in\R}\Pr\{\bx=x\|\rh\}\rh_{\{\bx=x\}}.
\eeq
If the measurement is a projective measurement of a discrete observable $A$,
this amounts to a mapping on the space $\tc(\cH)$ of trace-class operators on
$\cH$ such that 
\beq
T(\rh)=\sum_{x\in\R}E^{A}(\{x\})\rh E^{A}(\{x\}).
\eeq
Nakamura and Umegaki \cite{NU62} pointed out 
the analogy between quantum state reductions and the notion of
conditional expectation in probability theory
by showing that the dual map $T^*$ 
of $T$ is a normal norm-one projection, called a conditional expectation
\cite{Ume54,Tak79}, from the algebra $\cL(\cH)$ of bounded operators on $\cH$
to the commutant $\{A\}'$ of $A$, where $\{A\}'=
\{E^{A}(\De)\mid \De\in\cB(\bR)\}'$ if $A$ is unbounded.
They suggested that the state change caused by a measurement 
can be represented by such a conditional expectation from $\cL(\cH)$ to
$\{A\}'$ even if the observable $A$ has continuous spectrum.
However,  Arveson \cite{Arv67} showed that such a conditional expectation 
does not exist if $A$ has a continuous spectrum.
Based on the above results, Davies and Lewis \cite{DL70} proposed
a general framework for considering all the physically possible state changes
caused by measurements by abandoning the repeatability hypothesis as
the primary principle.
A {\it Davies-Lewis (DL) instrument} for a Hilbert space $\cH$ is 
a measure $\cI$ on the $\si$-field $\cB(\R)$ of Borel subsets of
the real line $\R$ with values in positive linear maps 
on the space  $\tc(\cH)$ of trace-class operators on $\cH$, countably additive
in the strong operator topology (i.e., $\cI(\bigcup_{n} \De_n)\rh
=\sum_{n}\cI(\De_n)\rh$ for all $\rh\in\tc(\cH)$ and 
disjoint sequence $\{\De_n\}$ in $\cB(\R)$), and normalized 
so that $\cI(\R)$ is trace-preserving (i.e., $\Tr[\cI(\R)\rh]=\Tr[\rh]$ for all
$\rh\in\tc(\cH)$).
A simple example of a DL instrument is given by a state change caused by
a projective measurement of a discrete observable $A$ by
\beq
\cI(\De)\rh=\sum_{x\in\De}E^{A}(\{x\})\rh E^{A}(\{x\})
\eeq
for all $\De\in\cB(\R), \rh\in\tc(\cH)$. 

Let $\bS$ be a system described by a Hilbert space $\cH$.
Consider a physically realizable measuring apparatus and 
denote it by $\bA(\bx)$. 
Here, $\bx$ represents the output variable of this apparatus and 
we assume it is real valued.
In quantum mechanics we cannot predict the value of the outcome
of each measurement and we can only deal with its statistical 
properties.
The statistical properties of the apparatus $\bA(\bx)$ 
are determined by (i) the probability distribution $\Pr\{\bx\in\De\|\rh\}$
of the outcome $\bx$ in an arbitrary state $\rh$, and
(ii) the state $\rh_{\{\bx\in\De\}}$ just after the measurement
 under the condition that the outcome satisfies $\bx\in\De$,
 where $\rh_{\{\bx\in\De\}}$ is defined for all $\De\in\cB(\R)$
 with $\Pr\{\bx\in\De\|\rh\}>0$, and it represents an indefinite
 state otherwise.
Thus, we assume the following postulate.
\begin{postulate}[Statistical properties of apparatuses]
\label{P:SPA}
To every apparatus $\bA(\bx)$  for $\cH$ uniquely associated are
a probability measure $\rh\mapsto\Pr\{\bx\in\De\|\rh\}$ for 
any $\rh\in\cS(\cH)$ and a density operator $\rh_{\{\bx\in\De\}}$ 
for any $\rh\in\cS(\cH)$ and $\De\in\cB(\R)$ with  $\Pr\{\bx\in\De\|\rh\}>0$.
\end{postulate} 
 
The proposal of Davies and Lewis can be stated as follows.

\begin{postulate}[Davies-Lewis thesis]
\label{P:DL}
For every apparatus $\bA(\bx)$ with output variable $\bx$ 
there exist  a unique DL instrument
$\cI$ satisfying
\beqa
\Pr\{\bx\in\De\|\rh\}&=&\Tr[\cI(\De)\rh],
\label{eq:DL1}\\
\rh_{\{\bx\in\De\}}&=&
\frac{\cI(\De)\rh}{\Tr[\cI(\De)\rh]}.
\label{eq:DL2}
\eeqa
\end{postulate}

In what follows,
we shall discuss a justification of the Davies-Lewis thesis following \cite{04URN}.
Let $\bA(\bx)$ and $\bA(\by)$ be two measuring apparatuses with the output
variables $\bx$ and $\by$, respectively.
Consider the successive measurements by $\bA(\bx)$ and $\bA(\by)$, carried
out first by $\bA(\bx)$ for the system $\bS$ in a state $\rh$
immediately followed by $\bA(\by)$ for the same system $\bS$.
Then, the joint probability distribution $\Pr\{\bx\in\De,\by\in\Ga\|\rh\}$ of
$\bx$ and $\by$ is given by
\beql{101121}
\Pr\{\bx\in\De,\by\in\Ga\|\rh\}=
\Pr\{\by\in\Ga\|\rh_{\{\bx\in\De\}}\}
\Pr\{\bx\in\De\|\rh\}.
\eeq
It is natural to assume that the above joint probability distribution 
has the following property.

\begin{postulate}[Mixing law for joint output probability]
\label{P:ML}
For any successive measurements carried out by apparatuses
$\bA(\bx)$ and $\bA(\by)$ in this order, the joint probability distribution
$\Pr\{\bx\in\De,\by\in\Ga\|\rh\}$ of output variables $\bx$ and $\by$
is an affine function in $\rh$.
\end{postulate}

This assumption can be justified as follows.
Since if the system $\bS$ is in
a state $\rh_1$ with probability $p$ and 
in a state $\rh_2$ with probability $1-p$, then
the joint probability distribution is given by 
$$
P=p\Pr\{\bx\in\De,\by\in\Ga\|\rh_1\}+
(1-p)\Pr\{\bx\in\De,\by\in\Ga\|\rh_2\}.
$$
On the other hand, in this case the system $\bS$ is 
in the state $\rh=p\rh_1+(1-p)\rh_2$, and hence the same
probability is given also by
$$
P=\Pr\{\bx\in\De,\by\in\Ga\|\rh\}.
$$
This concludes the mixing law for joint output probability.

From the postulate for statistical properties of apparatuses,
to any $\De\in\cB(\R)$ and $\rh\in\tc(\cH)$ corresponds
a unique trace-class operator 
\beq
\cI(\De,\rh)=
 \Pr\{\bx\in\De\|\rh\}\rh_{\{\bx\in\De\}}.
\eeq
Suppose, in particular, that $\bA(\by)$ is a measuring apparatus for
a measurement of a projection $E$.  Then, by
\Eq{101121} and \Eq{Born}, we have
\beqas
\Pr\{\bx\in\De,\by\in\{1\}\|\rh\}&=&
\Pr\{\by\in\{1\}\|\rh_{\{\bx\in\De\}}\}
\Pr\{\bx\in\De\|\rh\}\\
&=&
\Tr[E\rh_{\{\bx\in\De\}}]\Pr\{\bx\in\De\|\rh\}\\
&=&
\Tr[E\Pr\{\bx\in\De\|\rh\}\rh_{\{\bx\in\De\}}].
\eeqas
Thus, we have
\beql{101121b}
\Pr\{\bx\in\De,\by\in\{1\}\|\rh\}=\Tr[E\cI(\De,\rh)].
\eeq
Since $E$ is arbitrary, 
the mapping $\rh\mapsto\cI(\De,\rh)$ is an affine mapping
from $\tc(\cH)$ to $\tc(\cH)$, so that it uniquely extends to
a positive linear map from $\tc(\cH)$ to $\tc(\cH)$.
The finite additivity of the function 
$\De\mapsto \cI(\De,\rh)$ follows from the countable additivity of 
$\De\mapsto\Pr\{\bx\in\De,\by\in\{1\}\|\rh\}$.
Let $\{\De_n\}$ be an increasing sequence in $\cB(\R)$
such that $\bigcup_{n}\De_n=\De$.
Then, for any $\rh\in\cS(\cH)$ we have 
\beqas
\lim_{n\to\infty}
\|\cI(\De,\rh)-\cI(\De_{n},\rh)\|_{\tc}&=&
\Tr[\cI(\De,\rh)]-\lim_{n\to\infty}\Tr[\cI(\De_{n},\rh)]\\
&=&
\Pr\{\bx\in\De\|\rh\}-\lim_{n\to\infty}\Pr\{\bx\in\De_{n}\|\rh\}
=0,
\eeqas
where $\|\cdots\|_{\tc}$ denotes the trace norm on $\tc(\cH)$.
Thus, the mapping $\De\mapsto \cI(\De,\rh)$ is countably additive
in trace norm for any $\rh\in\cS(\cH)$.  Since $\tc(\cH)$ is linearly
generated by $\cS(\cH)$, this is the case for every
$\rh\in\tc(\cH)$.  Letting $\De=\R$ and $E=1$ in \Eq{101121b}, 
we have $\Tr[\cI(\R,\rh)]=1$ for all $\rh\in\tc(\cH)$.
It follows that $\rh\mapsto\cI(\R,\rh)$ is trace-preserving.
Letting $\cI(\De)\rh=\cI(\De,\rh)$, we have a DL instrument
$\cI$ satisfying \eq{DL1} and  \eq{DL2} for the apparatus $\bA(\bx)$.
Therefore, we have shown that the Davies-Lewis thesis is 
a consequence of the mixing law for joint output probability.

\subsection{Individual quantum state reduction}

It is natural to assume that the output variable $\bx$ can be 
read out with arbitrary precision, so that each instance
of measurement has the output value $\bx=x$.
Let $\rh_{\{\bx=x\}}$ be the state of the system $\bS$
at the time just after the measurement on input state $\rh$
provided that the measurement yields the output 
value $\bx=x$.
If $\Pr\{\bx\in\{x\}\|\rh\}>0$, 
the state $\rh_{\{\bx=x\}}$ is determined by the relation
\begin{equation}
\rh_{\{\bx=x\}}=\rh_{\{\bx\in\{x\}\}}.
\end{equation}
However, the above relation determines no $\rh_{\{\bx=x\}}$
if the output probability is continuously distributed.  
In order to determine states $\rh_{\{\bx=x\}}$, 
the following mathematical notion was introduced in \cite{85CA}.
A family $\{\rh_{\{\bx=x\}}|\ x\in{\bf R}\}$ of states is called a {\it family of
posterior states} for a DL instrument $\cI$ and a {\it prior state}
$\rh$, if it satisfies the following conditions.

(i) The function $x\mapsto\rh_{\{\bx=x\}}$ is Borel measurable.

(ii) For any Borel set $\De$, we have
\begin{equation}
{\cI}(\Delta)\rh
=
\int_{\Delta}\rh_{\{\bx=x\}}\Tr[d\cI(x)\rh].
\label{eq:226a}
\end{equation}

The following theorem ensures the existence of a family of posterior states
 \cite{85CA}．

\begin{theorem}[Existence of posterior states]
For any DL instrument $\cI$ and
prior state $\rh$, there exists a family of posterior
states essentially unique with respect to the probability measure
$\Tr[\cI(\cdot)\rh]$.
\end{theorem}

We define the {\it individual quantum state reduction} to be the correspondence 
from the input state $\rh$ to the family $\{\rh_{\{\bx=x\}}|\ x\in{\bf R}\}$ 
of posterior states for DL instrument $\cI$ 
of $\bA(\bx)$ and prior state $\rh$.
For distinction, we shall call the previously defined quantum
state reduction $\rh\mapsto\rh_{\{\bx\in\De\}}$ the {\it collective
quantum state reduction}.

The operational meaning of 
the individual quantum state reduction is given as
follows. Suppose that a measurement using the apparatus $\bA(\bx)$ on
input state $\rh$  is immediately followed by a measurement using
another apparatus $\bA(\by)$. Then, the joint probability distribution
$\Pr\{\bx\in\De,\by\in\De'\|\rh\}$ of the output variables $\bx$ and $\by$
is given by
\beqa\label{eq:101129}
\Pr\{\bx\in\De,\by\in\Ga\|\rh\}=
\int_{\De}
\Pr\{\by\in\Ga\|\rh_{\{\bx=x\}}\}
\Pr\{\bx\in dx\|\rh\}.
\eeqa
Thus, $\Pr\{\by\in\Ga\|\rh_{\{\bx=x\}}\}$ is the conditional
probability distribution of the output variable $\by$ of the  $\bA(\by)$
measurement immediately following the $\bA(\bx)$ measurement carried 
out on the input state $\rh$ given that the  $\bA(\bx)$ measurement
leads to the outcome $\bx=x$. 

\subsection{Complete positivity}

Since the postulate for statistical properties of apparatuses
(Postulate \ref{P:SPA}) and 
the mixing law for joint output probability
(Postulate \ref{P:ML})
 are considered
to be universally valid, we can conclude that every physically
realizable apparatus has a DL instrument representing its
statistical properties (Postulate \ref{P:DL}).
Thus, the problem of mathematically characterizing all the physically
possible quantum measurements is reduced to the problem
what class of DL instruments really can be considered to 
arise, in principle, from a physically realizable process.

A linear map $T$ from a *-algebra  $\cA$ to a *-algebra $\cB$ is called
{\it completely positive} if $T\otimes\id_{n}:
\cA\otimes M_{n}\mapsto \cB\otimes M_{n}$
is a positive map for every finite number $n$, where $M_n$
is the matrix algebra of order $n$ and $\id_{n}$ is the identity map on $M_n$.
The above condition is equivalent to requiring the relation
\beqa
\sum_{i,j=1}^{n}B_{i}T(A_{i}A^{\da}_{j})B^{\da}_{j}\ge 0
\eeqa
for any finite sequences  $A_{1},\ldots,A_{n}\in \cA$ and
$B_{1},\ldots,B_{n}\in\cB$.

Let $\cH$ be a Hilbert space.
A contractive completely positive map on the space $\tc(\cH)$
of trace-class operators is called an {\it operation} for $\cH$.
The dual map $T^*:\cL(\cH)\to\cL(\cH)$ of a completely positive 
map $T:\tc(\cH)\to\tc(\cH)$ is defined by $\Tr[T^*(A)\rh]=\Tr[AT(\rh)]$ 
for any $A\in\cL(\cH)$ and  $\rh\in\tc(\cH)$. 
This is a normal completely positive map on $\cL(\cH)$.
A DL instrument for $\cH$ is called a {\it completely positive (CP)
instrument},  or simply an {\it instrument}, if $\cI(\De)$ is
completely positive for every $\De\in\cB(\R)$.
 
Just like different reference frames may describe the same physical
process, different mathematical models may describe
the same measuring process.
For instance, an apparatus measuring an observable $A$ of
the system described by the Hilbert space $\cH$ is also
considered an apparatus for measuring the observable $A
\otimes 1$ of the system described by the Hilbert space
$\cH\otimes\cH'$ with Hilbert space $\cH'$ describing
another system. 
The above consideration naturally leads to the following postulate.

\begin{postulate}[Trivial extendability principle]
{\it For any apparatus $\bA(\bx)$ measuring a system $\bS$
and any quantum system $\bS'$ not interacting with $\bA(\bx)$ 
nor $\bS$,  there exists an apparatus $\bA(\bx')$ measuring system
$\bS+\bS'$  with the following statistical properties:
\beqa\label{eq:ex-1}
\Pr\{\bx'\in\De\|\rh\otimes\rh'\}
&=&
\Pr\{\bx\in\De\|\rh\},\\
(\rh\otimes\rh')_{\{\bx'\in\De\}}&=&\rh_{\{\bx\in\De\}}\otimes
\rh',
\label{eq:ex-2}
\eeqa
for any Borel set $\De$, any state $\rh$ of $\bS$, 
and any state $\rh'$ of $\bS'$.}
\end{postulate}

Now, suppose that $\bA(\bx)$ is an apparatus measuring
a system $\bS$ described by Hilbert space $\cH$,
and let $\cI$ be the DL instrument corresponding to $\bA(\bx)$.
Then, according to the above postulate
the physically identical measuring process can be described mathematically
by another apparatus $\bA(\bx')$ measuring the system $\bS+\bS'$
with another system $\bS'$ but without interacting with $\bS'$.
Let $\cI'$ be the DL instrument corresponding to $\bA(\bx')$.
Then, we have
\beqa
\cI'(\De)=\cI(\De)\otimes \id,
\eeqa
for all $\De\in\cB(\cH)$,
where $\id$ is the identity map on $\tc(\cH')$.
We say that an  DL instrument $\cI$ has the {\it trivial
extendability} if $\cI(\De)\otimes \id$ defines another
instrument for an arbitrary Hilbert space $\cH'$.
Thus, according to the trivial extendability postulate,
$\cI(\De)\otimes \id$ is required to be a positive map.
This means that the DL instrument $\cI$ should be a CP instrument.
Thus, the trivial extendability postulate leads to the
following postulate \cite{04URN}．

\begin{postulate}
For any apparatus $\bA(\bx)$ the corresponding instrument 
$\cI$ is completely positive.
\end{postulate}

\sloppy
Now, we have shown that the set of postulates 
\{Postulate 3,  Postulate 5,  Postulate 6\} is equivalent to
the set \{Postulate 3,  Postulate 4,  Postulate 7\}. 

The transpose map on the matrix algebra is 
a typical example of positive maps that are not completely positive.
The transpose map on $\tc(\cH)$ relative to an orthonormal basis
$\{\ph_n\}$ of $\cH$ is a bounded linear map
on $\tc(\cH)$ defined by
\beq
T(\ket{\ph_n}\bra{\ph_m})=\ket{\ph_m}\bra{\ph_n}
\eeq
for all $n,m$.
This is a trace-preserving positive map on $\tc(\cH)$, but not
completely positive.
For any observable $A=\sum_n n\ket{\ph_n}\bra{\ph_n}$, 
we have a DL instrument $\cI$ defined by 
\beqas
\cI(\De)\rh=\sum_{n\in\De}T(E^{A}(\{n\})\rh E^{A}(\{n\})).
\eeqas
According to the Davies-Lewis thesis this DL instrument should
correspond to the following measurement statisitics: 
\beqas
\Pr\{\bx=n\|\rh\}&=&\Tr[E^{A}(\{n\})\rh],\\
\rh_{\{\bx=n\}}&=&\frac{T(E^{A}(\{n\})\rh E^{A}(\{n\}))}
{\Tr[E^{A}(\{n\})\rh]}.
\eeqas
However,  according to the trivial extendability postulate
we can conclude that we have no measuring apparatus that 
physically realizes the above measurement statistics.

From the above, we conclude that physically realizable
measurement statistics is necessarily described by a CP instrument.
We say that two measuring apparatuses are {\it statistically 
equivalent} if they have the same statistical properties.
Our main objective is to determine the set of statistical equivalence classes
of physically realizable measurements.
Since every statistical equivalence class of physically realizable measurements
uniquely corresponds to a CP instrument, 
the problem is reduced to the problem as to
which CP instrument is physically realizable.
The purpose of the following argument is to show that
every CP instrument can be considered, in principle, to be physically 
realizable.

\subsection{Measuring processes}

Von Neumann \cite{vN32} showed that the projection postulate is
consistent with axioms of quantum mechanics.
Though von Neumann actually discussed the repeatability 
hypothesis for discrete observables with non-degenerate spectrum,
his argument can be easily generalized to the projection postulate
for discrete observables not necessarily with non-degenerate spectrum.
The process of a measurement always includes the interaction between
the object and the apparatus, and after the interaction the outcome 
of the measurement is obtained by measuring the meter in the apparatus.
Since the latter process can be done without the interaction between
the object and the apparatus,
the process of the measurement can be divided, at least, into the
above two stages.
Von Neumann showed that the statistical properties of the projective
measurement can be obtained by such a description of the measuring
process with an appropriate choice of the interaction, and 
the consistency of the projection postulate 
with quantum mechanics follows.

By generalizing von Neumann's argument, the standard models of
measuring processes were introduced in \cite{84QC}.
According to that formulation, a {\em measuring process} for 
(the system described by) a Hilbert space $\cH$
is defined as a quadruple $(\cK,\rh_0,U,M)$ consisting of a Hilbert space
$\cK$, a density operator $\rh_0$, a unitary operator $U$ on the tensor
produce Hilbert space $\cH\otimes\cK$, and a self-adjoint operator  $M$
on $\cK$.
A measuring process $(\cK,\rh_0,U,M)$ is said to be {\it pure}
if $\rh_0$ is a pure state, and it is said to be {\it separable} if
$\cK$ is separable.

The measuring process $(\cK,\rh_0,U,M)$ mathematically 
models the following description of a measurement.
The measurement is carried out by the interaction, referred to
as the {\it measuring interaction}, between
the {\it object} system $\bS$ and the {\it probe} system $\bP$, 
a part of the measuring apparatus $\bA(\bx)$ that actually takes part in
the interaction with the object $\bS$.
The probe system $\bP$ is described by the Hilbert space $\cK$
and prepared in the state $\rh_0$ just before the measurement.
The time evolution of the composite system $\bP+\bS$ 
during the measuring interaction is 
represented by the unitary operator $U$.
The outcome of the measurement is obtained by measuring 
the observable $M$ called the {\it meter observable} in the probe $\bP$.

Suppose that the measurement carried out by an apparatus 
$\bA(\bx)$ is described by a measuring process
$(\cK,\rh_0,U,M)$.
Then it follows from Axioms 1 to 4 that the statistical properties of
the apparatus $\bA(\bx)$ is given by 
\beqas
\Pr\{\bx\in\De\|\rh\}
&=&\Tr \left[\left(1\otimes E^{M}(\De)\right)
U(\rh\otimes\rh_0)U^{\da}\right],\\
\rh_{\{\bx\in\De\}}
&=&\frac{\Tr_{\cK}\left[\left(1\otimes E^{M}(\De)\right)
U(\rh\otimes\rh_0)U^{\da}\right]}
{\Tr \left[\left(1\otimes E^{M}(\De)\right)
U(\rh\otimes\rh_0)U^{\da}\right]},
\eeqas
where $\Tr_{\cK}$ stands for the partial trace on the Hilbert space
$\cK$; see \cite{84QC} for the detailed justification. 
Thus, if the measurement by the apparatus  $\bA(\bx)$ is described
by the measuring process $(\cK,\rh_0,U,M)$, the apparatus
$\bA(\bx)$ indeed has the instrument  $\cI$ determined by 
\beqa\label{eq:instrument_MP}
\cI(\De)\rh=\Tr_{\cK}\left[\left(1\otimes E^{M}(\De)\right)
U(\rh\otimes\rh_0)U^{\da}\right].
\eeqa
In this case, we call $\cI$ the instrument of the measuring process 
$(\cK,\rh_0,U,M)$.
Here, it is important to note that we never appeal to the projection 
postulate in order to derive the above relation \cite{84QC}.
In fact, \Eq{instrument_MP} holds even in the case where
the measurement of the meter-observable $M$ is not a projective
measurement; for a detailed discussion on this point see 
\cite{84QC,89RS,97QQ,97OQ,98QS,00MN,01OD}．

Now, we have shown that if the apparatus $\bA(\bx)$
is described by the measuring process $(\cK,\rh_0,U,M)$,
the statistical properties of $\bA(\bx)$ are determined by 
the instrument $\cI$ specified by \Eq{instrument_MP}.
Then, the problem is whether the converse is true.
The following theorem solves this problem \cite{83CR,84QC}.

\begin{theorem}[Realization theorem for instruments]
For any instrument $\cI$ for a Hilbert space $\cH$,
there exists a pure measuring process $(\cK,\rh_0,U,M)$ for $\cH$
such that $\cI$ is the instrument for $(\cK,\rh_0,U,M)$.
If $\cH$ is separable, $\cK$ can be made separable. 
\end{theorem}

From the above theorem, we conclude the following.
If we are given a physical measuring apparatus, 
that apparatus is considered to have its own statistical properties,
which are mathematically described by a DL instrument from the mixing law of
the joint output probability.
On the other hand, a mathematical description of 
a physical measuring apparatus should satisfy the 
trivial extendability, so that the DL instrument must
be a CP instrument.
It is a difficult problem to generally consider all 
the physically realizable measuring processes,
but for our purpose it suffices to consider a special
class of measuring processes, which we consider
as physically realizable and call ``measuring 
processes'' with a rigorous mathematical definition.
What is concluded by the realization theorem of instruments
is that for every physically realizable measuring apparatus $\bA(\bx)$,
there exists at least one measuring apparatus $\bA(\bx')$ in the above
class that is statistically equivalent to $\bA(\bx)$.
Therefore, it is concluded that a universal or an existential statement on
all the physically realizable measurements is justified 
if it is valid over the measurements carried out by measuring apparatuses
in that class as long as the statement concerns only statistical properties 
of measurements.  This gives an important approach to establishing 
the universally valid uncertainty principle. 

Now, we have justified the general measurement axiom formulated 
as follows.

\begin{axiom}[General measurement axiom]
\label{ax:GMA}
To every apparatus $\bA(\bx)$
for the system $\bS$ with the state
space $\cH$, there corresponds an instrument $\cI$ such that
the probability of the outcome $\bx\in\De$, where $\De\in\cB(\R)$, 
of the measurement in a state $\rh\in\cS(\cH)$ is given by 
\beqa
\Pr\{\bx\in\De\|\rh\}=\Tr[\bX(\De)\rh],
\eeqa
and if $\Pr\{\bx\in\De\|\rh\}>0$
the state $\rh_{\{\bx\in\De\}}$ just after the measurement 
under the condition that 
the measurement leads to the outcome $\bx\in \De$
is given by
\beqa
\rh_{\{\bx\in\De\}}=\frac{\bX(\De)\rh}{\Tr[\bX(\De)\rh]}.
\eeqa
Conversely, to every instrument $\cI$ there exists at least one 
apparatus $\bA(\bx)$ with the above statistical properties. 
\end{axiom}

A {\it probability operator-valued measure (POVM)} for a Hilbert space $\cH$ is 
a measure $\Pi$ on $\cB(\R)$ with values in positive operators 
on $\cH$, countably additive in strong operator topology, i.e., $\Pi(\bigcup_{n} \De_n\ps)
=\sum_{n}\Pi(\De_n)\ps$ for all $\ps\in\cH$ and 
disjoint sequence $\{\De_n\}$ in $\cB(\R)$, and normalized 
so that $\Pi(\R)=1$.
Let $\bX$ be an instrument for $\cH$.
The dual map $\bX(\De)^*$ of $\bX(\De)$ is a normal completely
positive map on the space $\cL(\cH)$ of bonded operators on $\cH$.
The relation
\beql{130a}
\Pi(\De)=\bX(\De)^{*}1,
\eeq
where $\De\in\cB(\cR)$, defines a 
POVM $\Pi$, called the POVM of $\bX$.
Conversely, every POVM arises in this way.
From Axiom \ref{ax:GMA}, 
Axiom \ref{ax:BSF} can be generalized as follows.

\begin{axiom}[Generalized statistical formula]
\label{ax:GSF}
To every apparatus $\bA(\bx)$ 
 for the system $\bS$ with the state
space $\cH$, there corresponds a POVM $\cI$ such that
the probability of the outcome $\bx\in\De$, where $\De\in\cB(\R)$, 
of the 
measurement in a state $\rh\in\cS(\cH)$ is given by 
\beqa
\Pr\{\bx\in\De\|\rh\}=\Tr[\Pi(\De)\rh].
\eeqa
Conversely, to every POVM $\Pi$ there exists at least one 
apparatus $\bA(\bx)$ with the above probability of the outcome. 
\end{axiom}

An apparatus $\bA(\bx)$ is said to {\it measure an observable} $A$ 
if its POVM is the spectral measure of $A$. 
Axiom \ref{ax:BSF} is derived from Axiom \ref{ax:GSF} under the
additional condition $\Pi=E^{A}$.
Let $A$ be a discrete observable.
The relation
\beqa
\bX^{A}(\De)\rh=\sum_{x\in \De}E^{A}(\{x\})\rh E^{A}(\{x\}),
\eeqa
where $\rh\in\tc(\cH)$, defines an instrument $\bX^{A}$,
called the instrument of the projective measurement of $A$.
In this case, the POVM of $\bX^{A}$ is $E^{A}$, and 
the projection postulate is derived from Axiom \ref{ax:GMA} under
the additional condition 
$\bX=\bX^{A}$.

The Wigner formula is generalized to the following.

\begin{theorem}[Generalized Wigner's formula]
Let $\bX_{1},\ldots,\bX_{n}$ be instruments for the system with
the state space $\cH$  in a state $\rh$ at time 0.
If one carries out measurements described by $\bX_{1},\ldots,\bX_{n}$  
at times $(0<)t_{1}<\cdots<t_{n}$ and otherwise leaves the system $\bS$ isolated,
then the joint probability distribution of the outcomes 
$\bx_{1},\ldots,\bx_{n}$ of those measurements is given by
\beqa
\lefteqn{\Pr\{\bx_{1}\in\De_{1},\bx_{2}\in\De_{2}\ldots,\bx_{n}\in\De_{n}\|\rh\}}
\qquad\qquad\nn\\
&=&
\Tr[\bX_{n}(\De_{n})\al(t_{n}-t_{n-1})\cdots
\bX_{2}(\De_2)\al(t_{2}-t_{1})\bX_{1}(\De)\al(t_{1})\rh],
\eeqa
where $\al$ is defined by 
$\al(t)\rh=e^{-iHt/\hbar}\rh e^{iHt/\hbar}$ for all 
$t\in\R$ and $\rh\in\c(\cH)$. 
\end{theorem}

Foundations of quantum measurement theory
based on the notion of instruments have
been developed in  
\cite{83CR,84QC,85CA,85CC,86IQ,88MR,93CA,95MM}.

\section{Uncertainty principle}

\subsection{Heisenberg's proof}
In 1927,  by considering the famous thought experiment of the $\gamma$
ray microscope, Heisenberg \cite{Hei27} showed the relation
\beqa\label{eq:Heisenberg}
\ep(Q) \et(P)\sim \hbar
\eeqa
for the measurement error $\ep(Q)$ of a position measurement
and the disturbance  $\et(P)$ of the momentum caused by that measurement.
He further stated that this is a straightforward mathematical consequence of 
the canonical commutation relation  $[Q,P]=i\hbar$ , and attempted to
give a formal proof based on the Dirac-Jordan theory.
In that proof he used the fact that the product of the spread of the position
and the spread of the momentum in a Gaussian wave function 
amounts to the Planck constant.
Immediately afterward, Kennard  \cite{Ken27} reformulated  
this relation in terms of the standard deviations $\sigma(Q)$ and $\sigma(P)$ 
of the position and the momentum, respectively, as 
\beqa\label{eq:Kennard}
\sigma(Q)\sigma(P)\ge\frac{\hbar}{2},
\eeqa
which he proved in any state $\ps$. 
In 1929 Robertson \cite{Rob29} further generalized and proved
this relation
to arbitrary pairs of observables $A$ and $B$ as 
\beqa\label{eq:Robertson}
\sigma(A)\sigma(B)\ge\frac{1}{2}|\bracket{\ps,[A,B]\ps}|.
\eeqa

Since then, most text books have shown the derivation of 
Robertson's relation \Eq{Robertson} in terms of 
the Schwarz inequality and then explained its physical meaning 
to be the quantitative relation such that if one measures the position
more precisely, then the momentum is more disturbed as in the
$\gamma$ ray thought experiment.

However, it is obvious that neither 
the Kennard inequality nor the Robertson inequality 
expresses the relation between the measurement error and
the disturbance, since the notion of standard deviation
has nothing to do with the properties of measuring apparatuses
but is determined solely by the state of the measured object.
In fact, Heisenberg's proof of \Eq{Heisenberg} runs as follows.
Heisenberg assumes that the measurement of the position 
with the error $\ep(Q)$ leaves the object in a state $\ps$ with
the standard deviation $\si(Q)$ satisfying $\sigma(Q)=\ep(Q)$.
Then, he uses the relation \Eq{Kennard} to obtain 
$\ep(Q)\si(P)\sim\hbar$, and concludes that it is because
the disturbance $\eta P$ satisfies \eq{Heisenberg}
that a measurement with small $\ep(Q)$ always increases
the standard deviation $\si(P)$ of the momentum.

The assumption used here is not correct that the measurement of the position 
with the error $\ep(Q)$ leaves the object in a state $\ps$ with
the standard deviation $\si(Q)$ satisfying $\sigma(Q)=\ep(Q)$.
This was revealed in the 1980s through the debate over the problem as to whether
there exists a detection limit derived from the uncertainty principle. 
In the rest of this section, we discuss this debate and the correct formulation of the   
uncertainty principle. 

\subsection{Gravitational wave detection and the uncertainty principle}

In the 1970s, from a simple quantum mechanical analysis on the performance of 
gravitational wave detectors it was generally accepted that a theoretical limit, called the
standard quantum limit (SQL), of the sensitivity of gravitational wave detectors
is derived from the uncertainty principle, and in particular that the SQL can be escapable 
by resonator type detectors but not escapable by non-resonator type detectors \cite{BV74,CTDSZ80}. 
However, in the 1980s a dispute arose among theorists on the validity of the SQL
\cite{Yue83,Cav85,88MS,89RS}．

A typical non-resonator type detector is an apparatus that estimates the existence or the strength 
of gravitational waves by detecting the change in the difference of the lengths 
of two orthogonal optical paths caused by the tidal force carried by the gravitational waves.
The measurement of the small change of the position of the mirror is assumed to obey quantum
mechanics.  Thus, the problem is how accurately one can predict the position of the mirror as
a free mass in the absence of gravitational waves.
If there is an inevitable error, the detectable force should give the displacement greater than the error,
and the gravitational waves cannot be detected if they are weaker than those which give such a 
minimum displacement.
 
In the standard argument \cite{BV74,CTDSZ80}, the time $t=0$ is set as the instant just after 
the first measurement and the time $t=\tau$ is the instant of the time just before the second 
measurement.  Then, it is claimed that according to Kennard's inequality \eq{Kennard} applied
to the standard deviations $\si( \hx(0))$ and  $\si(  \hp(0))$ of the position and the momentum 
just after the first measurement,  
the variance of the position $\hx$ increases until the 
time $\tau $ of the second measurement as 
\beqa
\quad\quad \si( \hx(\tau ))^2
\ge \si( \hx(0))^2 + \si( \hp(0))^2\tau ^2/m^2 
\geq  2\si( \hx(0))\si( \hp(0))\tau /m
\geq  \frac{\hbar \tau}{m}.
\eeqa
From the above, we obtained the SQL
 \begin{equation}
 \si(\hx(\tau )) \geq \sqrt{\frac{\hbar \tau}{m}}. 
                        \label{2.1.3}
 \end{equation}
In this way, the SQL has been explained as a straightforward consequence of 
Kennard's inequality \eq{Kennard}.
 
 Now, we suppose that a constant classical force $f$ acts on a mass $m$ 
from time $t=0$ to $t=\tau$. 
If $\De f$ is the minimum detectable force, then we have
$
\De f\tau^2/2m \ge \De \hx_{{\rm SQL}},
$
since the displacement at 
the time $\tau$ caused by this force should be more than $\De \hx_{{\rm SQL}}$.
Thus, the standard quantum limit  (SQL) for the detection of a weak classical force 
is obtained as 
\beq
\De f_{{\rm SQL}} =\sqrt{\frac{4\hbar m}{\tau^3}}.
\eeq

In 1983, Yuen \cite{Yue83} pointed out a serious flaw in this standard argument.
Since the evolution of a free mass is given by 
\begin{equation}
\hat{x}(t) = \hat{x}(0) + \hat{p}(0)t/m                        \label{2.1.4}
\end{equation}
 the variance of $\hx$  at time $\tau $ is given by
\beqa
 \si( \hx(\tau ))^2\!&=&\!  \si( \hx(0))^2 + \si(\hp(0))^2\tau ^2/m^2\\
& &{}+ \langle \delta \hat{x}(0)\delta \hat{p}(0)
  + \delta \hat{p}(0)\delta \hat{x}(0)\rangle \tau /m,\nonumber
  \qquad
\label{2.1.5}
\eeqa
 where $\Delta \hat{x} = \hat{x} - \langle \hat{x}\rangle $ 
 and $\Delta x^2 = \langle \Delta \hat{x}^2\rangle $, etc. 
Thus the
standard argument implicitly assumes that the last term --- we
shall call it the {\em correlation term} --- in Eq.~(\ref{2.1.5})
is non-negative. Yuen's assertion \cite{Yue83} is that some
measurements of $\hx$ leave the free mass in a state with the
negative correlation term.

In other words, the measurement of the position of a free-mass
at $t=\tau$  has no uncertainty, if the state at $t=0$ is an eigenstate of
$\hx(\ta)$.
Any eigenstate of  $\hx(\ta)$ is not normalizable but 
there are (normalized) wave functions arbitrarily near it, 
and the contractive states are among them.

However, if the measurement is only approximately accurate,
namely, the measurement outcome at time  $\tau$ includes the additional 
error to the actual position $\hat{x}(\tau)$, then the expected uncertainty 
of the measurement outcome is considered to include the measurement error 
in addition to the quantum mechanical uncertainty.   
Thus, the problem is reduced to the problem as to whether it is possible to 
realize, in principle, the measurement such that its measurement error for 
the position $\hx(0)$ is negligibly small but the mass is left 
in a state arbitrarily near to an eigenstate of the observable $\hx(\ta)$. 

The existence of such a measurement contradicts the Heisenberg type inquality
\eq{Heisenberg}.
In fact, by \Eq{Heisenberg} we have
\beqa
\ep[\hx(0)]\et[\hx(\ta)]\ge\frac{\ta\hbar}{2m},
\eeqa
and hence if the measurement error is $\ep[\hx(0)]\approx0$, the
disturbance of $\hx(\ta)$ satisfies
$\et[\hx(\ta)]\sim\infty$, so that it is impossible to have
the relation
$\De \hx(\ta)\approx0$ 
in the state after the measurement.

A dispute arose as to whether such a measurement is possible or not,
and the theoretical aspect of the dispute was settled by the result \cite{88MS}
showing such a measurement can be carried out by a model that is obtained by
a straightforward modification of the von Neumann model \cite{vN32} of position measurement
\cite{Mad88}.

\subsection{Noise and disturbance in quantum measurement}
Let $(\cK,\rh_0,U,M)$ be a measuring process for a system $\bS$
described by a Hilbert space $\cH$.
For this measuring process and an observable $A$ of $\bS$,  
we define the {\it noise operator} $N(A)$,
and {\it disturbance operator} $D(A)$ by
\beqa
N(A)&=&U^{\da}(1\otimes M) U-A\otimes 1,\\
D(A)&=&U^{\da}(A\otimes 1) U-A\otimes 1.
\eeqa
Their means, $\bracket{N(A)}$ and
$\bracket{D(A)}$,
in the state  $\rh\otimes\rh_0$ are
called the {\it mean noise} and {\it mean disturbance},
respectively, for observable $A$ in a state $\rh$.
Their root-mean-squares (rms's), 
$\bracket{N(A)^2}^{1/2}$ and 
$\bracket{D(A)^2}^{1/2}$, 
in the state  $\rh\otimes\rh_0$ are
called the {\it (rms) noise} and {\it (rms) disturbance},
respectively,  for observable $A$ in a state $\rh$,
and denoted by $\ep(A)$ and $\et(A)$.

We also define {\it mean noise operator} $n(A)$ and
{\it mean disturbance operator} $d(A)$ by
\beqa
n(A)&=&\Tr_{\cK}[N(A)(1\otimes\rh_0)],\\
d(A)&=&\Tr_{\cK}[D(A)(1\otimes\rh_0)].
\eeqa

The $n$th moment operator  $\Pi^{(n)}$ of a POVM $\Pi$
is defined by 
\beqas
\bracket{\et,\Pi^{(n)}\xi}&=&\int_{\R}
x^n\bracket{\et,\Pi(dx)\xi}, \quad (\xi\in\dom(\Pi^{(n)}), \et\in\cH),
\\
\dom(\Pi^{(n)})&=&\{\xi\in\cH\mid
\int_{\R} x^{2n} \bracket{\xi,\Pi(dx)\xi}<\infty\}.
\eeqas

Let $T$ be an operation for $\cH$.
For any observable $A$,
denote by 
$T^{*}E^{A}$
the POVM defined by $(T^{*}E^{A})(\De)=T^*(E^{A}(\De))$.
If $A$ is bounded, it is easy to see that $T^{*}(A^n)$ is
the $n$th moment operator of $T^{*}E^{A}$.
If $A$ is unbounded, we define $T^{*}(A^n)$ as
the $n$th moment operator of $T^{*}E^{A}$,
i.e., 
$T^*(A^n)=(T^{*}E^{A})^{(n)}$.

The following theorem shows that the mean noise, 
the rms noise, and the mean noise operator are determined by
the POVM of the measuring process and  the mean disturbance, 
the rms disturbance, and the mean disturbance 
operator are determined by the operation of the measuring process.

\begin{theorem} 
Let $(\cK,\rh_0,U,M)$ be a measuring process for 
a Hilbert space $\cH$, and let $T$ and  $\Pi$ be the 
corresponding POVM and operation.
Then, we have 
\beqa
n(A)&=&\Pi^{(1)}-A,\\
d(A)&=&T^{*}(A)-A,\\
\bracket{N(A)}&=&
\Tr[\Pi^{(1)}\rh]-\Tr[A\rh],\\
\bracket{D(A)}&=&\Tr[AT(\rh)]-\Tr[A\rh],\\
\ep(A)^2&=& 
\Tr[\Pi^{(2)}\rh]-\Tr[\Pi^{(1)}\rh A]-\Tr[\Pi^{(1)}A\rh]
+\Tr[A^2\rh],\\
\et(A)^2&=&
\Tr[A^2T(\rh)]-\Tr[AT(\rh A)]-\Tr[AT(A\rh)]+\Tr[A^2\rh].
\eeqa
Here, we assume that $\rh$ satisfies $A\sqrt{\rh}\in\sic(\cH)$ 
and that the relevant traces are convergent.
\end{theorem}

Following the proposal introduced in Heisenberg \cite{Hei27},
we call the relation
\beqa\label{eq:HUP}
\ep(A)\et(B)\ge \frac{1}{2}|\Tr([A,B]\rh)|
\eeqa
the {\it Heisenberg type inequality}.

\subsection{Von Neumann's measurement}

Von Neumann \cite{vN32} introduced the following measuring process 
of a position measurement.
Caves  \cite{Cav85} showed that this measurement satisfies 
the SQL.
Here, we shall show that this measurement satisfies the Heisenberg 
inequality. 

The measured object $\bS$ is a one-dimensional quantum system
with position $\hat{x}$, momentum $\hat{p}_{x}$, 
satisfying $[x,p_x]=i\hbar$, and Hamiltonian
$H_{\bS}$.
Suppose that
the object $\bS$ interacts with the probe $\bP$ in the apparatus
$\bA(\bx)$
from time $t$ to $t+\De t$ and 
becomes free from time $t+\De t$.
In von Neumann's measuring process, the probe $\bP$ is 
a one-dimensional quantum system
with position $\hat{y}$, momentum $\hat{p}_{y}$, 
satisfying $[y,p_y]=i\hbar$, and Hamiltonian
$H_{\bP}$.
The meter observable in the probe $\bP$
is the position $\hat{y}$ of $\bP$.
The interaction between the object $\bS$ and the probe $\bP$ 
is given by
\beqa\label{eq:vN}
H_{\bS\bP}=\hat{x}\hat{p}_{y},
\eeqa
so that the total Hamiltonian of the composite system $\bS+\bP$
is given by
\beqa
H_{\bS+\bP}=H_{\bS}\otimes 1+ 1\otimes H_{\bP}+
KH_{\bS\bP},
\eeqa
where the coupling constant $K$ is so large that free 
Hamltonians can be neglected.
The time duration $\De t$ is assumed to satisfy $K\De t=1$.
Thus, the time evolution of the composite
system $\bS+\bP$ in the time duration $(t,t+\De t)$ is
given by 
\beq
U=e^{-i\hat{x}\hat{p}_{y}/\hbar}.
\eeq
Let $\xi$ be the initial state of the probe.
Then, the von Neumann model corresponds to the measuring process 
$(L^2(\bR),\xi,e^{-i\hat{x}\hat{p}_{y}/\hbar},\hat{y})$,
and its instrument is given by
\beqa
\cI(\De)\rh=\int_{\De}\xi(\hx-x1)\rh\xi(\hx-x1)^{\da}dx.
\eeqa
Solving Heisenberg's equation of motion, we have
\beqa
\hat{x}(t+\De t)&=&\hat{x}(t),\\
\hat{y}(t+\De t)&=&\hat{x}(t)+\hat{y}(t),\\
\hat{p}_{x}(t+\De t)&=&\hat{p}_{x}(t)-\hat{p}_{y}(t),\\
\hat{p_{y}}(t+\De t)&=&\hat{p}_{y}(t).
\eeqa
Thus, the noise operator and the disturbance operator are given by
\beqa
N(\hat{x})
&=&\hat{y}(t+\De t)-\hat{x}(t)=\hat{y}(t),\\
D(\hat{p}_{x})
&=&\hat{p}_{x}(t+\De t)-\hat{p}_{x}(t)=-\hat{p}_{y}(t).
\eeqa
Let $\si( \hat{y})$ and $\si( \hat{p}_{y})$ be 
the standard deviations of the position and the momentum of the probe,
respectively, at the time $t$ of the measurement.
Then, by the Kennard inequality, \eq{Kennard}, we have
\beqa
\ep(\hat{x})\et(\hat{p}_{x})\ge\si(\hat{y})\, \si(\hat{p}_{y})
&\ge&\frac{\hbar}{2}.
\eeqa
Thus, the Heisenberg type inequality
\eq{HUP} holds for von Neumann's measuring process
\cite{02KB5E}．

\subsection{Contractive state measurement}

The notion of contractive state measurements proposed 
by Yuen \cite{Yue83} has been shown to be
realized by the following measuring
process \cite{88MS,89RS,90QP,01CQSR}．

The measured object $\bS$, the probe $\bP$, and the time of 
interaction are described in the same way as von Neumann's model.
The interaction $H_{\bS\bP}$ is given by
\beql{829ox}
H_{\bS\bP}=\frac{K\pi}{3\sqrt{3}}
\{2(\hat{x}\hat{p}_{y}-\hat{p}_{x}\hat{y})
+(\hat{x}\hat{p}_{x}-\hat{y}\hat{p}_{y})\}.
\eeq
Thus, this model of measurement corresponds to
the measuring process
$$
(L^2(\bR),\xi,
\exp[ -i \frac{\pi}{3\sqrt{3}\hbar}
\{2(\hat{x}\hat{p}_{y}-\hat{p}_{x}\hat{y})
+(\hat{x}\hat{p}_{x}-\hat{y}\hat{p}_{y})\}],\hat{y}),
$$
and its instrument is given by
\beqa
\cI(\De)\rh=\int_{\De}e^{-ix\hat{p}_{x}/\hbar}\ketbra{\xi}e^{ix\hat{p}_{x}/\hbar}
\Tr[E^{\hx}(dx)\rh].
\eeqa

Solving Heisenberg's equation of motion, we have
\beqa
\hat{x}(t+\De t)&=&\hat{x}(t)-\hat{y}(t),\label{eq:ozawa-model-1}\\
\hat{y}(t+\De t)&=&\hat{x}(t),\label{eq:ozawa-model-2}\\
\hat{p}_{x}(t+\De t)&=&-\hat{p}_{y}(t),\label{eq:ozawa-model-3}\\
\hat{p}_{y}(t+\De t)
&=&\hat{p}_{x}(t)+\hat{p}_{y}(t)\label{eq:ozawa-model-4}.
\eeqa
Thus, the noise operator and the disturbance operator are given by
\beqa
N(\hat{x})
&=&\hat{y}(t+\De t)-\hat{x}(t)=0,\\
D(\hat{p}_{x})
&=&\hat{p}_{x}(t+\De t)-\hat{p}_{x}(t)=-\hat{p}_{y}(t)
-\hat{p}_{x}(t),
\eeqa
and hence
\beq
\ep(\hat{x})\et(\hat{p}_{x})=0.
\eeq
Thus, this model does not satisfy the Heisenberg type inequality, \eq{HUP}
\cite{02KB5E}．

\subsection{Universal uncertainty principle}

What relation between the error and the disturbance holds for arbitrary measurements?
The following theorem generally holds \cite{03UVR,03HUR,04URN}．

\begin{theorem}[Universal uncertainty principle]
The rms error $\ep(A)$, the rms disturbance $\et(B)$, and the standard deviations $\si(A), \si(B)$
satisfy the relation
\beqa
\ep(A)\et(B)+\ep(A)\si( B)+\si(A)\et(B)
\ge\frac{1}{2}|\Tr([A,B]\rh)|
\eeqa
for any observables $A,B$, state $\rh$, and instrument $\cI$.
\end{theorem}

\begin{theorem}[Condition for the Heisenberg type inequality]
The rms error $\ep(A)$ and the rms disturbance $\et(B)$ satisfy the relation
\beqa
\ep(A)\et(B)+
\frac{1}{2}|\Tr([n(A),B])+
\Tr([A,d(B)])|
\ge\frac{1}{2}|\Tr([A,B]\rh)|
\eeqa
for any observables $A,B$, state $\rh$, and instrument $\cI$.
Moreover, if the mean error $\bracket{N(A)}$ of $A$ and the mean disturbance
$\bracket{D(B)}$ of $B$ are independent of the object state, then the Heisenberg type
inequality {\rm \eq{HUP}} holds.
\end{theorem}
In fact, if $\bracket{N(A)}$ and $\bracket{D(B)}$ are independent of the object state,
then $n(A)$ and $d(B)$ are scalar operators, so that we have $[n(A),B]=[A,d(B)]=0$.

From the universal uncertainty principle, there are two typical cases where the Heisenberg
type inequality fails and we have a new trade-off relation in each case. 

(i) Constraint for error-free measurements: In the case where $\et(B)=0$,  the relation
\beqa\label{eq:nondisturbing}
\ep(A)\si(B)
\ge\frac{1}{2}|\Tr([A,B])|
\eeqa
holds for the error of $A$ and the standard deviation of $B$.

(ii) Constraint for non-disturbing measurements:
In the case where $\et(B)=0$, the relation
\beqa\label{eq:noiseless}
\si(A)\et(B)\ge\frac{1}{2}|\Tr([A,B])|
\eeqa
holds for the disturbance of $B$ and the standard deviation of $A$.

The model of the contractive state measurement \eq{829ox} is an instance of 
error-free measurements and reveals the possibility of a measurement 
breaking the standard quantum limit for gravitational wave detection.
In the next section, we shall show that the new constraint for non-disturbing 
measurements leads to a quantitative generalization of the Wigner-Araki-Yanase
theorem and an accuracy constraint for quantum computing. 

\section{Accuracy limits of quantum computing}

\subsection{Decoherence and conservation laws in quantum computing}

The prime factorization problem has been used for public key cryptography
such as the RSA protocol, 
since no efficient algorithm has been found for this problem. 
However, Shor  \cite{Sho94} found an efficient algorithm for quantum
computers solving prime factoring in 1994.
Since then, active researches have been developed as to the realizability of 
quantum computers.

A major part of the problem of realizability of a quantum computer is 
the problem of decoherence.
In general, decoherence in quantum computer components can be 
classified into two classes: (i) the environment induced decoherence,
arising from the interaction between the computer memory
and the environment, and (ii) the controller induced decoherence, 
arising from the interaction between the computer register 
and the control system of the quantum logic gate operation.
According to the theory of fault-tolerant quantum computing,
provided the noise in individual quantum gates is
below a certain threshold, it is possible to efficiently perform
arbitrarily large quantum computing \cite{NC00}.
The environment induced decoherence may be overcome by using materials 
with long decoherence time. 
On the other hand, the controller induced decoherence
poses a dilemma between controllability and decoherence;
the control needs coupling, whereas the coupling causes decoherence.
Thus, the problem is reduced to the problem as to whether the controller 
induced decoherence is derived to be inevitable from fundamental physical laws
and the problem of its quantitative evaluation.

One of the reasons why the controller induced decoherence is considered to be
inevitable in quantum state control is the existence of conservation laws in 
nature.
The Wigner-Araki-Yanase (WAY) theorem \cite{Wig52,AY60} is  
a starting point for the research on how conservation laws impede
quantum state control.
The WAY theorem states that no measuring interaction realizes
a measurement with absolute precision for an observable 
not commuting with additive conserved quantity.

\subsection{Quantitative generalization of the Wigner-Araki-Yanase theorem}

We show that the above new constraint on the accuracy of non-disturbing
measurements \eq{nondisturbing} can be used to derive the quantitative 
expression of the WAY theorem as follows \cite{02CLU,03UPQ}.

\begin{theorem}[Quantitative generalization of the WAY theorem]
\label{th:QGWAY}
For any measuring process $(\cK,\xi,U,M)$,
if observables $L_{1}$ and $L_{2}$ on Hilbert spaces $\cH$ and $\cK$,
respectively, satisfy $[U,L_{1}\otimes 1+1\otimes L_{2}]=0$ and $[M,L_{2}]=0$,
then for any observable $A$ on $\cH$ we have
\beqa\label{eq:WAY}
\ep(A)^{2}
\ge 
\frac
{|\bracket{[A,L_{1}]}|^{2}}
{4\si( L_{1})^{2}+4\si( L_{2})^{2}},
\eeqa
where the mean and standard deviations are taken for the initial states
of the system and the apparatus.
\end{theorem}
The proof runs as follows.  
By the relation $[U,L_{1}\otimes 1+1\otimes L_{2}]=0$,
the interaction between the system $\bS$ and the probe $\bP$
does not disturb $L_{1}\otimes 1+1\otimes L_{2}$.
Moreover, by the relation $[M,L_{2}]=0$, the subsequent measurement of
the probe observable $M$ can be done without disturbing 
$L_{1}\otimes 1+1\otimes L_{2}$.
Thus, a measuring process describing the same measurement, 
in which we regard $\bS+\bP$ as the measured object, satisfies 
$\et(L_{1}\otimes 1+1\otimes L_{2})=0$.
By substituting $B=L_{1}\otimes 1+1\otimes L_{2}$ in 
inequality \eq{nondisturbing}, we have 
\beqa
\ep(A)^{2}
\ge\frac{|\bracket{[A\otimes 1,L_{1}\otimes 1+1\otimes L_{2}]}|}
{4\si(L_{1}\otimes 1+1\otimes L_{2})^2},
\eeqa
and hence we have
\Eq{WAY} from the relations
$\bracket{[A\otimes 1,L_{1}\otimes 1+1\otimes L_{2}]}=
[A,L_{1}]$ and $\si(L_{1}\otimes 1+1\otimes L_{2})^{2}
=\si (L_{1})^{2}+\si( L_{2})^{2}$.

Yanase \cite{Yan61} derived  the accuracy limit for measurements of a 
spin component under the angular momentum conservation law.
Let $A$ be the $z$-component $S_{z}\otimes 1$ of a spin 1/2 particle $\bS$,
let $L_{1}$ be the $x$-component $S_{x}\otimes 1$ of $\bS$, 
and let $L_{2}$ be the $x$-component $1\otimes S_{x}$ of the probe $\bP$.
Yanase showed that the error probability $P_{e}$ satisfies
$ 
P_{e}\sim{\hbar^{2}}/{16\bracket{L_{2}^{2}}}.
$ 
In this case, we have
$|\bracket{[A,L_{1}]}|
=|\bracket{[S_{z},S_{x}]}|
=\hbar |\bracket{S_{y}}|
\le {\hbar^{2}}/{2}$,
and hence by Theorem \ref{th:QGWAY} we have
\beqa
\max_{\ps}\, \ep(A)^{2}
\ge \frac{\hbar^{4}}{4\hbar^{2}+16(\De L_{2})^{2}},
\eeqa
where $\max$ is taken over all the possible states $\ps$ of the object $\bS$.
From the relation $P_{e}=\ep(S_{z})^{2}/\hbar^{2}$, 
we have
\beqa\label{eq:spin-error}
\max_{\ps}\, P_{e}
\ge \frac{1}{4+16(\De L_{2}/\hbar)^{2}}.
\eeqa
Therefore,  inequality  \eq{WAY} for that case improves Yanase's result.

From  \Eq{spin-error}, it is concluded that the angular momentum conservation law
prevents the interaction for a precise spin measurement.
However, this result does not imply the unmeasurability of spin.  
It is clear from \Eq{WAY} that the inevitable error is inversely proportional
to the variance of the conserved quantity included in the apparatus.
An apparatus for high precision measurements is usually of 
macroscopic size and has a large amount of the conserved quantity,
and hence the practical apparatus circumvents the present limitation.
On the other hand, as discussed in the next section,
it is an interesting problem how an elementary quantum
logic gate in a small integrated circuit can operate with very high precision
demanded for fault-tolerant computing. 

\subsection{Quantum limits for the realization of quantum computing}

In the current paradigm
the strategies  for the realization of quantum computing
can be summarized as follows \cite{NC00}.

(1)  To physically represent computational qubits by spin components of spin 1/2 systems,
for the feasibility of initialization and read-out.

(2)  To physically realize elementary logic gates by1 qubit rotation operation and controlled
not (CNOT) operation between 2 qubits, and any quantum circuit can be built up from those
two sorts of unitary operations.
  
(3) To clear the accuracy threshold, every operation should be implemented with
the error probability below $10^{-5}-10^{-6}$. 

From the above it can be concluded that 
since rotations of the spin and the CNOT do not conserve the spin,
it has been shown from the above strategies
that if those gates are implemented by physical
interactions obeying the angular momentum conservation law,
then the unavoidable noise similar to the WAY theorem arises
\cite{02CQC}．
However, not every quantum gates will play the same role as the
measuring apparatus, and in fact there are quantum gate that
obey the angular momentum conservation law like the SWAP gate.
Thus, it is not always possible to estimate the error probability from 
inequality \eq{spin-error} quantifying the WAY theorem, but 
some useful arguments have been known for estimating the error probability 
for several gates \cite{02CQC,03UPQ,07CLI}．

Along this line, we have now established a method for estimating the error 
probability for arbitrary unitary gates.
Let $\bS$ be a spin 1/2 system described by a Hilbert space $\cH_{\bS}$,  
and let $\{\ket{0},\ket{1}\}$ be the eigenbasis of the $z$-component of the spin.
An arbitrary unitary gate $U_{\bS}$ on $\cH_{\bS}$ can be represented by 
$$
U_{\bS}=e^{i\ph}\left(\cos\frac{\theta}{2} 1+i\sin\frac{\theta}{2}\vec{n}\cdot
\vec{\sigma}\right)  
$$
with uniquely determined angles $\ph,\theta$ with $0\le\ph<2\pi$, $0\le \theta\le \pi$
and a unit vector $\vec{n}=(l_x,l_y,l_z)$, where $\vec{\sigma}=(\si_x,\si_y,\si_z)$ is
the vector consisting of the Pauli operators.
An implementation of $U_{\bS}$ is a pair $\al=(U,\ket{\xi})$ consisting of a unitary
operator $U$ of the composite system   $\bS+\bA$ with a system $\bA$, called the ancilla,
described by a Hilbert space $\cH_{\bA}$.
An implementation $\al=(U,\ket{\xi})$ defines a trace-preserving quantum operation $\cE_{\al}$  by
\beq
\cE_{\al}(\rh)=\Tr_{\bA}[U(\rh\otimes\ket{\xi}\bra{\xi})U^{\da}].
\eeq
for any density operator $\rh$ of the system $\bS$, where $\Tr_{\bA}$ stands 
for the partial trace over $\cH_{\bA}$.
On the other hand, the gate $U_{\bS}$ defines a unitary operation ${\rm ad}H$ by 
$
{\rm ad}U_{\bS}(\rh)=U_{\bS}\rh U_{\bS}^{\da}.
$
The {\em gate error probability} $P_e$ of the implementation $\al=(U,\ket{\xi})$
is defined as the completely bounded distance between $\cE_{\al}$ and ${\rm ad}U_{\bS}$,
i.e.,  
\beql{CB-distance}
D_{CB}(\cE_{\al},U_{\bS})={\rm sup}_{n,\rh}
D(\cE_{\al}\otimes {\rm id}_{n}(\rh),{\rm ad}H\otimes {\rm id}_{n}(\rh)),
\eeq
where $n$ runs over positive integers, ${\rm id}_{n}$  is the identity
operation on the matrix algebra  $M_n$,  
$\rh$ is a density operator on $\cH\otimes \C^{n}$,
and $D$ stands for the trace distance $D(\rh_1,\rh_2)=
\frac{1}{2}\Tr[|\rh_1-\rh_2|]$.
On the other hand the {\em gate fidelity} of the implementation  
$\al=(U,\ket{\xi})$ is defined by 
\beq
F(\cE_{\al},U_{\bS})
={\rm inf}_{\ket{\ps}}F(\ps),
\eeq
where $\ket{\ps}$ varies over the state vectors  of $\bS$ and
$F(\ps)$ is the fidelity between the two states $\cE_{\al}(\ket{\ps}\bra{\ps})$
and ${\rm ad}U_{\bS}(\ket{\ps}\bra{\ps})$ given by
$
F(\ps)=
\bracket{\ps|U_{\bS}^{\da}\cE_{\al}(\ket{\ps}\bra{\ps})U_{\bS}|\ps}^{1/2}.
$
The above measures of imperfection of the implementation $\al=(U,\ket{\xi})$
satisfy the relation \cite{NC00}
\beq
1-F(\cE_{\al},U_{\bS} )^{2}\le D_{CB}(\cE_{\al},U_{\bS}).
\eeq
The left-hand side is called the {\em gate infidelity} of the implementation
$\al=(U,\ket{\xi})$. 
Now, we assume that the implementation $(U,\ket{\xi})$ is rotationally 
invariant; namely, it satisfies the spin conservation law for 
$j=x,y,z$ components,  
$
[U,S_{j}\otimes 1+1\otimes L_{j}]=0,
$
and that the spin quantum number of the ancilla is $N/2$.
Then, a lower bound of the gate infidelity is given as follows \cite{09GFA}．
If $0\le\theta\le\pi/2$, we have
\beqa
\frac{\sin^2\theta}{4+4N^2}\le 1-F(\cE_{\al},U_{\bS} )^{2},
\eeqa
and if $\pi/2\le\theta\le \pi$, we have
\beqa
\frac{1}{4+4N^2}\le 1-F(\cE_{\al},U_{\bS} )^{2}.
\eeqa

\section{Interpretation of quantum theory}

\subsection{Simultaneous measurements of non-commuting 
observables}

It has long been accepted that two observables are simultaneously
measurable if and only if their corresponding operators commute.
However, this is true only when we take it as the statement that
two observables are simultaneously measurable {\em in any state} 
if and only if their corresponding operators commute.
In fact, in the singlet state of a system consisting of two spin-1/2 
particles any two components of the spin of the first particle 
is simultaneously measurable.
In order to do so, we have only to measure one component indirectly 
through the measurement of the same component of the second particle,
which is strictly anti-correlated with the same component of the 
first particle, and to measure the other component directly at the same time.

In what follows we present a mathematical theory of simultaneous measurability,
and give a theoretical basis for simultaneous measurability of non-commuting
observables.

We say that two observables $A,B$ are {\em commuting} in a state $\rh$ if
for any Borel sets $\De,\Ga$ we have $[E^A(\De),E^B(\Ga)]\rh=0$.
In this case, the {\em joint probability distribution}  $\mu$ of observables
 $A,B$ in the state $\rh$ is defined by
\beqa
\mu^{A,B}_{\rh}(\De\times\Ga)=\Tr[E^A(\De)E^B(\Ga)\rh].
\eeqa
We say that two observables $A,B$ have a {\em quantum identical correlation}
in a state $\rh$, and write $A\equiv_\rh B$,
if they are commuting in $\rh$ and the joint probability distribution
satisfies
\beqa
\mu^{A,B}_{\rh}(\{(x,y)\in\R^2\mid x=y\})=1.
\eeqa
In this case, two observables $A,B$ are considered to be simultaneously
measurable in $\rh$ and their measurement outcomes are always identical.

Let $f,g$ be Borel functions.
A measuring process $(\cK,\xi,U,M)$ for a Hilbert space $\cH$ 
is said to {\em simultaneously measure} observables $A,B$ 
with $f,g$ in a state $\rh$ if we have 
\beqa
U^{\da}(1\otimes f(M))U &\equiv_{\rh\otimes\ketbra{\xi}}&A\otimes 1,\\
U^{\da}(1\otimes g(M))U &\equiv_{\rh\otimes\ketbra{\xi}}&B\otimes 1.
\eeqa
Two observables $A,B$ are said to be {\em simultaneously measurable} in a state 
$\rh$ if there is a measuring process $(\cK,\xi,U,M)$ 
together with Borel functions $f,g$ such that $(\cK,\xi,U,M)$ 
simultaneously measures observables $A,B$ with $f,g$ in a state $\rh$.
From the following theorem, the notion of simultaneous measurement is determined
by the POVM of a measuring process \cite{06QPC}．
\begin{theorem}
A measuring process $(\cK,\xi,U,M)$ for a Hilbert space $\cH$ 
simultaneously measures observables $A,B$ 
with Borel functions $f,g$ in a state $\rh$ if and only if the POVM
$\Pi$ of the measuring process $(\cK,\xi,U,M)$ satisfies 
\beqa
\Tr[\Pi(f^{-1}(\De))E^{A}(\Ga)\rh]=
\Tr[\Pi(g^{-1}(\De))E^{A}(\Ga)\rh]=0
\eeqa
for every disjoint Borel subsets $\De,\Ga$.
\end{theorem}

Let $C(A_1,A_2,\rh)$ be the projection 
onto the minimum invariant subspace of $\cH$ of $A_1$ and $A_2$
including the range of $\rh$.
Let $C(A_1,\rh)=C(A_1,I,\rh)$.
The conceptual difference between the commutativity and simultaneous
measurability is given by the following theorems \cite{07SMN};
see also M. Ozawa, {\em Quantum reality and measurement: 
A quantum logical approach}, Found.\ Phys.\ {\bf 41}  (2011), 592--607．

\begin{theorem}
Two observables $A,B$ are commuting in a state $\rh$ if and only if 
there exists a POVM $\Pi$ on $\R^2$ such that for every Borel subset $\De$ 
we have 
\beqa
\Pi(\De\times\R)C(A,B,\rh)&=&E^{A}(\De)C(A,B,\rh),\\
\Pi(\R\times \De)C(A,B,\rh)&=&E^{B}(\De)C(A,B,\rh).
\eeqa
\end{theorem}

\begin{theorem}
Two observables $A,B$ are simultaneously measurable in a state $\rh$ if and only if 
there exists a POVM $\Pi$ on $\R^2$ such that for every Borel subset $\De$ 
we have 
\beqa
\Pi(\De\times\R)C(A,\rh)&=&E^{A}(\De)C(A,\rh),\\
\Pi(\R\times \De)C(B,\rh)&=&E^{B}(\De)C(B,\rh).
\eeqa
\end{theorem}

\subsection{Quantum reality and quantum set theory}

Let $\bS$ be a quantum system described by a Hilbert space $\cH$.
For any observable $A$ of $\bS$ and an interval $\De$,
we denote by $A\in \De$ the proposition that 
the value of the observable $A$ is in the interval $\De$,
and call it an {\em atomic observational proposition}. 
{\em Observational propositions} are those constructed from
atomic observational propositions using logical symbols
of negation $\Not$, conjunction $\And$, disjunction $\Or$, and implication $\Then$.
The lattice  $\cQ$ of projections on the Hilbert space $\cH$ is called
the {\em quantum logic} of the system $\bS$; symbols $\And$, $\Or$, and $\perp$ denote
meet, join, and orthogonal complement, respectively. 
We define the {\em $\cQ$-valued truth value} $\val{\ph}$ of an observational proposition $\ph$
by the following rules.

(i) $\val{A\in \De}=E^{A}(\De);$

(ii) $\val{\Not \ph}=\val{\ph}^{\perp};$

(iii) $\val{\ph_1\And \ph_2}=\val{\ph_1}\And\val{\ph_2};$

(iv) $\val{\ph_1\Or \ph_2}=\val{\ph_1}\Or\val{\ph_2};$

(v) $\val{\ph_1\Then\ph_2}=
\val{\ph_1}^{\perp}\Or(\val{\ph_1}\And\val{\ph_2})$.

\noindent
Then, the Born statistical formula can be extended to the following 
relation:
\beqa
\Pr\{A_1\in \De_1,\ldots,A_n\in \De_n\| \rh\}
=
\Tr[\val{A_1\in \De_1\And \cdots\And A_n\in \De_n}\rh].
\eeqa

However, by this method we cannot determine the truth value or 
the probability of some observational proposition such as $A=B$, meaning 
that the value of the observable $A$ and the observable $B$ are identical.
In the recent investigation  \cite{07TPQ}, it becomes clear that 
quantum set theory is quite useful for such a problem on extending 
the probability interpretation of quantum mechanics;
in fact, the notion of quantum identical correlations between two observables, 
which plays an important role in the theory of quantum measurements as mentioned
in the preceding subsection, has been shown to be equivalent with the
notion of equality between two real numbers in quantum set theory, and
hence that notion has acquired a natural and independent motivation.  

In 1963 P. J. Cohen proved that the continuum hypothesis is independent from
the axioms of ZFC set theory by inventing a new method, called forcing, 
to construct a new model of ZFC.
In 1966 Scott and Solovay reformulated forcing by the method of 
Boolean-valued models of set theory, which was eventually widely accepted as
a tractable approach to Cohen's forcing. 
In 1981 G. Takeuti \cite{Ta81} introduced quantum set theory by extending 
the construction of Boolean-valued models from Boolean logic to quantum logic.

In what follows, we survey quantum set theory based on the recent development
\cite{07TPQ}; see also M. Ozawa,
{\em Orthomodular-valued models for quantum set theory}, 
arXiv:0908.0367.
Let $\cQ$ be a complete orthomodular lattice, in which
the orthogonal complementation $\perp$ corresponds to negation,
the infimun operation $\And$ corresponds to disjunction,
and the supremum operation  $\Or$ corresponds to conjunction.
Although the operation $\Then$ corresponding to implication is ambiguous in general, 
here we define $a\Then b=a^{\perp}\Or(a\And b)$ for all $a,b\in\cQ$;
the operation $\Then$ so defined is often called the Sasaki arrow.
The $\cQ$-valued universe $\V^{(\cQ)}$ of set theory is defined by a transfinite 
recursion on subclasses $\V_{\al}^{(\cQ)}$ as follows, where ``${\rm On}$'' 
stands for the class of ordinal numbers.

(i) $V_{0}^{(\cQ)}=\emptyset$;

(ii)  $V_{\al+1}^{(\cQ)}=\{u|\   u:\dom(u)\to\cQ,\
\dom(u)\subseteq V_{\al}^{(\cQ)}\}$;

(iii)  For limit ordinal $\al$,
$V_{\al}^{(\cQ)}=\bigcup_{\be<\al}V_{\be}^{(\cQ)}$;

(iv) $V^{(\cQ)}=\bigcup_{\al\in{\rm On}}V_{\al}^{(\cQ)}$.

\noindent
If $\cQ$ is a complete Boolean algebra $\cB$, 
the model $V^{(\cQ)}$ coincides with the Scott-Solovay Boolean-valued 
model $V^{(\cB)}$.
If $\cQ$ is the projection lattice on a Hilbert space,  $V^{(\cQ)}$ coincides with Takeuti's
model.
If $\cQ={\bf 2}(=\{0,1\})$, this reduces to the usual interpretation of set theory
in the two-valued logic.

An element of $V^{(\cQ)}$ is called a {\em $\cQ$-valued set}.
From the above definition, $\cQ$-valued set $u$ is a function on the
set $\dom(u)$, a subset consisting of the $\cQ$-valued sets in some 
$V_{\al}^{(\cQ)}$, with values in $\cQ$, 
and $u(x)$ essentially represents the truth value in $\cQ$ of 
the relation $x\in u$ with an appropriate modification, if necessary.
For any $\cQ$-valued sets $u,\ v$, the truth values of atomic propositions
$u=v$ and $u\in v$ are defined as follows:

(i) $\val{u=v}=\inf_{u'\in\dom(u)}(u(u')\Then\val{u'\in
v})\And\inf_{v'\in\dom(v)}(v(v')\Then\val{v'\in
u})$;

(ii)
$\val{u\in
v}=\sup_{v'\in\dom(v)}(v(v')\And\val{u=v'})$.

Any well-formed formula $\ph$ is constructed from
atomic propositions and logical symbols
$\Not$,
$\And$, $\Or$, $\Then$, 
$(\forall x\in y)$, $(\exists x\in y)$, and 
$(\forall x)$, $(\exists x)$, 
by well-known composition rules.
The quantifiers $(\forall x\in y)$ and  $(\exists x\in y)$ are called bounded
quantifiers and the quantifiers $(\forall x)$ and $(\exists x)$ are called
unbounded quantifiers. 
Any formula without unbounded quantifiers is called a bounded formula.
The truth value of a statement $\phi$ is defined as follows:

(i) $\val{\Not \phi}=\val{\phi}^{\perp}$;

(ii) $\val{\phi_{1}\And \phi_{2}}=\val{\phi_{1}}\And\val{\phi_{2}}$;

(iii) $\val{\phi_{1}\Or \phi_{2}}=\val{\phi_{1}}\Or\val{\phi_{2}}$;

(iv) $\val{\phi_{1}\Then \phi_{2}}=\val{\phi_{1}}\Then\val{\phi_{2}}$;

(v) $\val{(\forall    u'\in u)\phi(u')}=\inf_{u'\in
\dom(u)}\val{\phi(u')}$;

(vi)  $\val{(\exists     u'\in u)\phi(u')}=\sup_{u'\in
\dom(u)}\val{\phi(u')}$;

(vii) $\val{(\forall    x)\phi(x)}=\inf_{u\in
V^{(\cQ)}}\val{\phi(u)}$;

(viii) $\val{(\exists     x)\phi(x)}=\sup_{u\in
V^{(\cQ)}}\val{\phi(u)}$.

Let $V$ be the universe of ZFC set theory.
For any $a\in V$, the $\cQ$-valued set 
$\check{a}$ is defined by 
$\dom(\check{a})=\{\check{x}|\ x\in a\}$ and 
$\check{a}(\check{x})=1$ for all $x\in a$.
The correspondence $a\mapsto\check{a}$ embeds 
the universe $V$ into the $\cQ$-valued universe $V^{(\cQ)}$.
Then, the relation between sets $a,b$ is equivalent to
the relation between $\cQ$-valued sets $\check{a},\check{b}$; namely,
$a\in   b$，$a\not\in b$，$a=b$，$a\not=b$ are equivalent to 
$\val{\check{a}\in\check{ b}}=1$，
$\val{\check{a}\in\check{ b}}=0$，$\val{\check{a}=\check{b}}=1$，
$\val{\check{a}=\check{b}}=0$, respectively.

It is an important problem to investigate what statements hold in the $\cQ$-valued 
universe.
If  $\cQ$ is a complete Boolean algebra $\cB$, it is well known that 
the following transfer principle holds: for any formula $\ph(x_1,\ldots,x_n)$ 
provable in ZFC, we have 
$$
\val{\ph(u_1,\ldots,u_n)}=1
$$ 
for every $u_1,\ldots,u_n\in \VB$.
If $\cQ$ is not distributive, the above transfer principle does not hold in general.
For instance, neither the transitivity of equality nor the substitution law for equality
hold.
However, we can see that the $\cQ$-valued universe has a rich structure, since
it includes many Boolean-valued universes as subuniverses.

A subset $\cS$ of $\cQ$ is called a commuting system if every two elements 
of $\cS$ commute.
For any $\cQ$-valued sets $u_1,\ldots,u_n$, let $L(u_1,\ldots,u_n)$ be 
the subset of $\cQ$ consisting of all elements of $\cQ$ which are used to
construct $u_1,\ldots,u_n$.  
Let $\cuniv(u_1,\ldots,u_n)$ be the maximum element $p\in\cQ$ such that
$p$ commutes with all elements of  $L(u_1,\ldots,u_n)$ and 
$p\And L(u_1,\ldots,u_n)$ is a commuting system.
Then, for any complete orthomodular lattice $\cQ$, the following transfer
principle holds \cite{07TPQ}: {\em for any bounded formula $\ph(x_1,\ldots,x_n)$ 
provable in ZFC, we have  
$$
\val{\ph(u_1,\ldots,u_n)}\ge\cuniv(u_1,\ldots,u_n)
$$ 
for every $u_1,\ldots,u_n\in \VB$.}

It can be seen that the set of natural numbers in $V^{(\cQ)}$ is $\check{\om}$,
and that the set of rational numbers in  $V^{(\cQ)}$ is $\check{\Q}$.
However, the set of real numbers in  $V^{(\cQ)}$ does not necessarily 
correspond to $\check{\R}$.
Here,  the set of real numbers in  $V^{(\cQ)}$ is defined as the $\cQ$-valued
set consisting of the Dedekind cuts of $\check{\Q}$ in $V^{(\cQ)}$.
Then, we have $\val{\check{\R}\subseteq\R_{\cQ}}=1$.

Suppose that $\cQ$ is the quantum logic of the system $\bS$ 
described by the Hilbert space $\cH$, namely, the lattice of projections on $\cH$.
Then,  for every $u$ such that $\val{u\in\R_{\cQ}}=1$ the projections $E_{\la}$
with $\la\in\R$ defined by $E_{\la}=\val{u\le\check{\la}}$ form a resolution
of the identity, and hence  $u$ corresponds to the self-adjoint
operator $\hat{u}$ defined by
$$
\hat{u}=\int_{\R}\la\,dE_{\la}.
$$
The above relation sets up a one-to-one correspondence between the real numbers in 
$V^{(\cQ)}$ and the observables of the quantum system $\bS$.
Let $\tilde{A}$ be the real number in $V^\cQ$ corresponding to an
observable $A$ and let  $\tilde{\De}$ be the interval in $V^\cQ$
corresponding to an interval $\De$ in the real line.
Then, we have
\beqa
\val{\tilde{A}\in\tilde{\De}}=E^{A}(\De).
\eeqa
Thus, {\em quantum observables are nothing but real numbers in quantum set theory,
and  observational propositions can be embedded in propositions on the real numbers 
in quantum set theory without changing their truth values \cite{07TPQ}}．

In quantum set theory the truth value of the equality relation has been defined.
Using this, for any observables $A,B$ we can determine the truth value of 
the observational proposition $A=B$.  Namely, we define
\beqa
\val{A=B}=\val{\tA=\tB}.
\eeqa
Then, it can be seen that this equality relation is equivalent to the quantum 
identical correlation.
In fact, we can see that $A\equiv_\rh B$ holds for a state $\rh$ if and only if
$\Tr[\val{\tA=\tB}\rh]=1$, or equivalently $\val{\tA=\tB}$ coincides with the
projection onto the subspace generated by vector states $\ps$ for which
the relation $A\equiv_\ps B$ holds \cite{07TPQ}．
Thus, {\em the notion of the identical correlation between two observables is 
nothing but the equality relation between two reals in quantum set theory.}

As above, quantum set theory is a useful way to systematically extend
the interpretation of quantum mechanics.
We can expect that quantum set theory will play an important role 
in describing a consistent image of quantum reality, which has been
a long-standing mystery in modern physics.

%
%
%
\providecommand{\bysame}{\leavevmode\hbox to3em{\hrulefill}\thinspace}

\end{document}